\documentclass[12pt]{iopart}
\usepackage{epsfig}
\usepackage{fontenc}
\usepackage{amssymb}
\usepackage{graphicx,subfigure}

\newcommand{\nc}{\newcommand}

\nc{\be}[1]{\begin{equation} \mbox{$\label{#1}$}}
\nc{\bea}[1]{\begin{eqnarray} \mbox{$\label{#1}$}}
\nc{\Section}[2]{\section{#2}\label{#1}}
\nc{\Bibitem}[1]{\bibitem{#1}}
\nc{\Label}[1]{\label{#1}}

\nc{\w}{\omega}
\nc{\eea}{\end{eqnarray}}
\nc{\ee}{\end{equation}}

\def\ltsima{$\; \buildrel < \over \sim \;$}
\def\simlt{\lower.5ex\hbox{\ltsima}}
\def\gtsima{$\; \buildrel > \over \sim \;$}
\def\simgt{\lower.5ex\hbox{\gtsima}}

\newcommand\rhom{\rho_{\rm m}}
\newcommand\rhor{\rho_{\rm r}}
\newcommand\deltam{\delta_{\rm m}}
\newcommand\omegam{\Omega_{\rm m}}

\nc{\lcdm}{{$\Lambda$CDM} }
\nc{\rar}{\rightarrow}

\begin{document}

\title[]
{Combined constraints on Cardassian models
from supernovae, CMB and large-scale structure observations}
\author{
M Amarzguioui$^1$\footnote[1]{morad@astro.uio.no}, 
\O\ Elgar\o y$^{1,2}$\footnote[2]{oelgaroy@astro.uio.no}
and T Multam\"{a}ki$^2$\footnote[3]{tuomas@nordita.dk}
}
\address{
$^1$Institute of theoretical astrophysics, University of Oslo, Box 1029, 0315 Oslo, Norway\\
$^2$NORDITA, Blegdamsvej 17, DK-2100 Copenhagen, Denmark
}

\date{\today}

\begin{abstract}

We confront Cardassian models with recent observational data. 
These models can be viewed either as purely phenomenological modifications 
of the Friedmann equation, or as arising from cosmic 
fluids with non-standard properties. 
In the first case, we find that the models are consistent with the data 
for a wide range of parameters but no significant preference over the \lcdm
model is found. In the latter case we find that 
the fact that the sound speed is non-zero in these models makes them 
inconsistent with the galaxy power spectrum from the Sloan Digital Sky 
Survey.    
\end{abstract}

\section{Introduction}

The Universe is expanding at an accelerating rate.  We have direct 
evidence for this from the observed magnitude-redshift relationship of 
supernovae type Ia (SNIa) \cite{riess, perlmutter}.
Cosmic acceleration 
requires a contribution to the energy density with negative pressure, 
the simplest possibility being a cosmological constant. Independent 
evidence for a non-standard contribution to the energy budget of 
the universe comes from e.g. the combination of the power spectrum 
of the cosmic microwave 
background (CMB) temperature anisotropies and large-scale 
structure (LSS):  on the one hand, the position of the 
first peak in the CMB is consistent 
with the universe having zero spatial curvature, which means that 
the energy density is equal to the critical density. On the other hand, 
observations of large-scale structure, e.g. the clustering of galaxies, 
show that the contribution of standard 
sources of energy density, whether luminous or dark, is only a 
fraction of the critical density.  Thus, an extra, unknown component 
is needed to explain the observations \cite{efstathiou, tegmark}.

The model resulting from reintroducing Einstein's cosmological constant,  
$\Lambda$, in a universe with baryons, radiation, and cold dark 
matter (CDM) is consistent with all large-scale cosmological observations 
like the anisotropies in the CMB radiation 
and the power spectrum of galaxies \cite{sdss}.
However, the modern interpretation of $\Lambda$ is that it represents 
the energy of the vacuum, and from the viewpoint of quantum field 
theory, the tiny value inferred for it from observations is highly 
unnatural \cite{weinberg}.  Faced with this problem, 
the popular choice is to set $\Lambda$ to zero and invoke a new component 
to explain the acceleration.  This does not 
solve the problem of the smallness of $\Lambda$; if it is indeed equal 
to zero, one still needs to understand the physical mechanism behind this.  
However, one may hope that $\Lambda=0$ 
may be easier to explain than a small, but nonzero $\Lambda$.  
The question is then what the unknown component driving the accelerated 
phase of expansion is.  One needs to introduce a component with negative 
pressure, and this can be done e.g. by invoking a slowly evolving scalar 
field 
\cite{wetterich, peebles, ratra}
, or 
a negative-pressure fluid, e.g. a Chaplygin gas \cite{kamenshchik, bilic}
A negative-pressure fluid, however, can 
be problematic due to its fluctuations. Fluctuations of the unknown
fluid can grow very rapidly and hence LSS surveys can place strict
constraints on such models (see e.g. \cite{bean, sandvik}). 
In order to circumvent this, it is sometimes assumed that the new fluid 
does not fluctuate on the scales of interest and its existence is 
visible only through modified background evolution. 

As an alternative to fluid models of dark energy, modifications 
of the Friedmann equation have been considered by several authors 
\cite{dim1, dim2,dim3, dvaliturner, carroll, card, card2}.  
In an earlier paper \cite{elgar1}, we investigated a model \cite{dim1,dvaliturner}
where the left-hand side 
of the Friedmann equation is modified.  In the present paper, we 
will look at the so-called Cardassian models 
\cite{card,card2}, where the right-hand side of the Friedmann equation is 
modified.

The structure of this paper is as follows.  In section 2 we summarize the 
basic properties of the Cardassian models.  In section 3 we
consider fits to SNIa and WMAP data.  In section 4 we extend the analysis to include 
probes of the large-scale distribution of matter, and in section 5 
we summarize our constraints.  Finally, section 6 contains our conclusions.   

\section{Cardassian models}

Cardassian models were introduced in \cite{card} as a possible alternative
to explain the acceleration of the universe by a model that has no 
energy components in addition to ordinary matter. Necessarily, the Friedmann 
equation must be modified:
\be{origmodel}
H^2={8\pi G\over 3}(\rhom+\rhor)+A\rhom^n,
\ee
where $\rhom$ is the energy density of matter (baryonic and dark), $\rhor$ 
is the energy density of radiation radiation and $n,\, A$ are parameters 
of the model. The requirement that universe undergoes acceleration at late times 
sets $n<2/3$. Phenomenologically there is no lower limit but conventionally
only models with $n>0$ are considered.
The appearance of the extra term on the right hand side was originally motivated by 
brane cosmology \cite{chung}. This approach was, however criticized in \cite{cline}
where it was shown that in the scenario of \cite{chung}, the weak energy condition
was violated on the brane. We will ignore the radiation term in the following 
discussion for simplicity 
but keep it in the CMB calculations as there it will have an effect. For 
SNIa constraints, the radiation contribution is negligible.

More recently, the Cardassian scenario has been generalized (see e.g. 
\cite{card2}) and its connection to fundamental physics has become less
apparent. Instead, one can view the Cardassian models as a class of 
phenomenological models
that attempt to explain the accelerated expansion of the universe without 
explicitly resorting to adding new cosmic fluids, such as a cosmological 
constant, to the Friedmann equation.

A most general formulation of the Cardassian framework is to consider Friedmann equations
that depend only on the matter density in the universe, thus
\be{gencard}
H^2={8\pi G\over 3}g(\rhom),
\ee
where $\rhom$ is the energy density of matter (CDM+baryons).
Nucleosynthesis dictates that at early times, the evolution
of the universe must be very close to the standard $H^2\sim \rho$, hence
$g(\rhom)\sim \rhom$ for large $\rhom$. In principle, one can consider arbitrary
functions $g$ or expansions such as 
\be{gexpan}
g(\rhom)=\rhom+\sum_{i=0}^\infty c_i\rhom^{-i},
\ee
where $c_i$ are parameters. The original Cardassian scenario then corresponds
to considering just the leading term in the expansion.

A commonly considered model within the Cardassian framework is the 
Modified Polytropic Cardassian (MPC) model \cite{card2}, 
which extends the original Cardassian proposal by introducing an
additional parameter. In the MPC model,
\be{mpc}
g(\rhom)=g^{MPC}(\rhom)\equiv\rhom\Big(1+({\rhom\over\rho_{\rm C}})^{q(n-1)}\Big)^{1/q},
\ee
where $q$ is the new parameter. The original scenario is reached in the special
case $q=1$. 
Specializing to a flat universe, and defining $\omegam\equiv 8\pi G \rho_{\rm m}^0/3 H_0^2$
(where $\rhom^0$ is the present day value),
the MPC model can be written as
\be{mpc2}
\Big({H\over H_0}\Big)^2=
\omegam a^{-3}\Big(1+\Omega_Ca^{3q(1-n)}\Big)^{1/q},
\ee
where $\Omega_C\equiv (\omegam^{-q}-1)$.
Similarly, the generalized model, Eq. \ref{gencard}, can be written
as
\be{gencard2}
\Big({H\over H_0}\Big)^2=f(a),
\ee
with $f(a)$ in the MPC case given by Eq. (\ref{mpc2}).

The main observational constraints to any non-standard scenario are SNIa, CMB and LSS.
Which probe is most effective depends on the details of the considered scenario.
If the model in question contains new, fluctuating fluids, the gravitational
clustering can be radically modified \cite{tegmark} and can lead to very tight constraints.
If, on the other hand, only the background evolution is modified via the
Friedmann equation, the growth of large scale structure, while modified,
may not differ too radically \cite{multam,lue}.

The Cardassian scenario has two possible physical interpretations: the first and 
maybe most common one is to consider the modified Friedmann equation arising
from modified gravity/extra dimensional constructions, leading to effective 
terms on the RHS of the equation. Another alternative \cite{card2}, the
so-called fluid interpretation, is to consider the modified Friedmann equation 
as a result of having an additional exotic fluid in the universe.

Here, we consider both possibilities using data from SNIa, CMB and
LSS observations to see how well the Cardassian scenarios are constrained by
the combination of current data. First we study the effective model with no
fluctuations in the Cardassian fluid and later consider the effects
of the fluctuations.

\section{Fits to CMB and SNIa data}

\subsection{Modified background evolution}

When fitting the CMB data with the Cardassian models, 
one should in principle start from the full underlying theory.  
This can be problematic.  If e.g. the Cardassian models arise from 
a scenario with extra dimensions, studying the evolution of perturbations 
is difficult because of the bulk-brane interactions and 
hence we will in the following use a simplified approach where we solve 
the standard 4-d perturbed Boltzmann and fluid equations, but with 
the background evolution given by Eq. (\ref{origmodel}) (also see discussion
in \cite{deff3}).

For convenience, we will parameterize the effect of the extra term 
in the Friedmann equation by a dark energy component with an 
effective equation of state $w(a)$. As long as the extra component
does not fluctuate, i.e. it only has an effect on the background
evolution, such a parameterization is equivalent to modifying
the Friedmann equation.

For a general Friedmann equation, Eq. (\ref{gencard}), we can write the
expansion of the universe in terms of an effective dark energy fluid.
A dark energy component with equation of state 
$w(a)$ has a density which varies with the scale factor $a$ according to 
\begin{equation}
\rho_{\rm D}(a) = \rho_{0\rm D} \exp\left\{-3\int_1^{a} \frac{da'}{a'}[1+w(a')]\right\}, 
\label{eq:eq2}
\end{equation}
so for a flat universe, $\Omega_{\rm D}=1-\Omega_{\rm m}$, the 
standard Friedmann equation is 
\begin{equation}
\left(\frac{H}{H_0}\right)^2 = {\Omega_{\rm m}\over a^3} 
+(1-\Omega_{\rm m})\exp\left\{-3\int_1^{a} \frac{da'}{a'}[1+w(a')]\right\}.
\label{eq:eq3}
\end{equation}

\begin{figure}
\begin{center}
{\centering
\mbox
{\psfig{figure=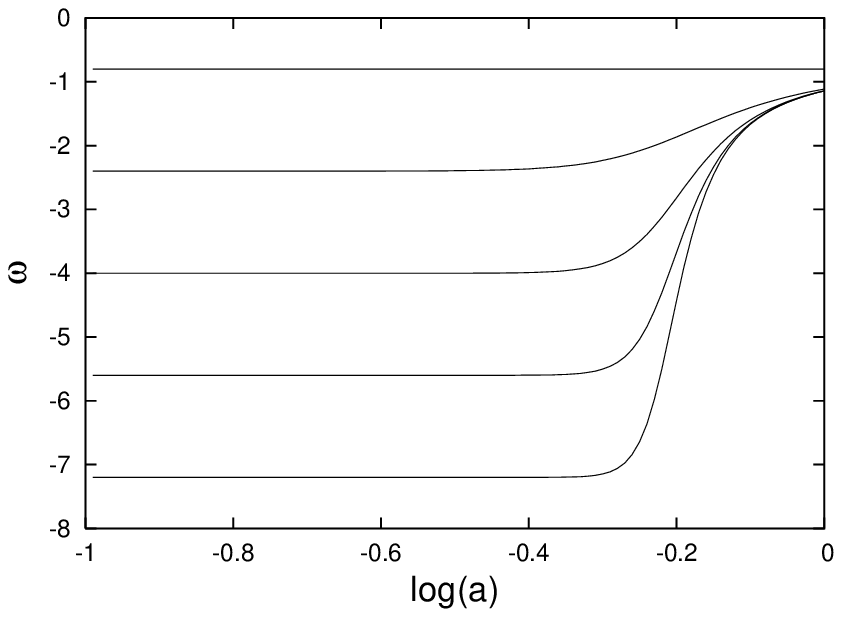,height=5cm,width=5cm}}
\mbox
{\psfig{figure=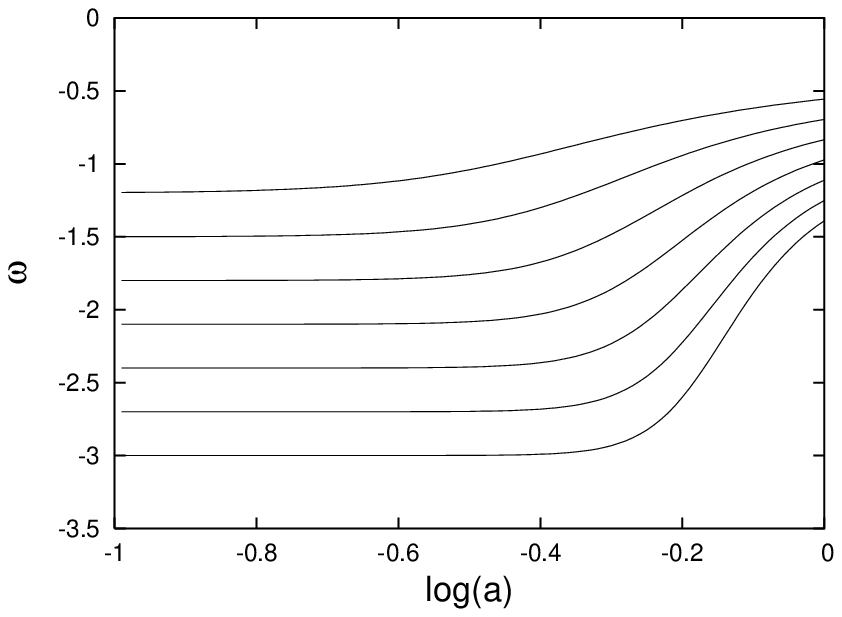,height=5cm,width=5cm}}
}
\caption{The equation of state $w(a)$ as a function of $\log(a)$,
for $\Omega_{\rm m}=0.3$. Left panel: $n=0.2,\ q=1\,({\rm top}),\ 3,\ 5,\ 7,\ 9\,({\rm bottom})$, right panel: $q=3,\ n=0.0\,({\rm bottom}),\ 0.1,\ ...,\ 0.6\,({\rm top})$.}
\label{fig:fig1}
\end{center}
\end{figure}

Using this along with Eq. (\ref{gencard2}), it straightforward to see that
the effective equation of state is (for $f(a)\neq \omegam a^{-3}$)

\be{efluid}
w(a) = -1-{a\over 3} {f'(a)+3\omegam a^{-4}\over f(a)-\omegam a^{-3}}.
\ee
For the special case $f(a)=\omegam a^{-3}+(1-\omegam)$ corresponding to the \lcdm model,
$w(a)=-1$ as expected. In the case of the MPC model, the equation of state
of the effective dark energy fluid is
\be{mpcwa}
w(a) = (1-n){a^{3(1-n)q}\Omega_C \Big(1+\Omega_Ca^{3q(1-n)}\Big)^{1/q-1}\over
1-\Big(1+\Omega_C a^{3(1-n)q}\Big)^{1/q}}.
\ee
At early times, $a\rar 0$, the equation of state tends to a constant:
$w\rar q(n-1)$. Also note that for the special case $q=1$, the equation
of state is constant $w=n-1$. For $q\neq 1$, $w$ evolves in such a way
that $w$ is more negative in the past and increases at late times.
The equation of state of the effective fluid in the MPC model
is shown in Fig. \ref{fig:fig1} for different values of the parameters.

\subsection{SNIa and CMB data}

For the supernova fit we use the sample of 194 SNIa presented in \cite{barris}.
We fit three parameters to to the data: $\Omega_{\rm m}$, $n$
and $q$ and values considered are $0 < \Omega_{\rm m} < 1$, $0 < n < 2/3$
and $q\geq 1$. The Hubble parameter $h$ is also involved in 
the fits, but is of little interest here, and we choose to 
marginalize over it (by maximizing the likelihood).

For fitting the CMB TT power spectrum \cite{wmap1} we use
the likelihood code provided by the WMAP team 
\cite{verde}\footnote{http:// lambda.gsfc.nasa.gov/}.  
The CMB power spectra is computed by using CMBFAST code 
\cite{cmbfast},  
version 4.5.1\footnote{http://www.cmbfast.org/}.
For each point in the parameter space $(\Omega_{\rm m},n,q,h)$,
we calculate the CMB TT power spectrum keeping the amplitude
of the fluctuations a free parameter by finding
the best fit amplitude for each set of parameters. In other words,
we only fit the shape and not the amplitude of the power
spectrum. In calculating the CMB power spectrum we use
$\Omega_{\rm b}=0.044$ (so that we vary the density of cold 
dark matter, $\Omega_{\rm c}$) 
and ignore reionization effects. We have checked that setting
the reionization optical depth to the WMAP value makes very little
difference in the resulting contours. Fixing 
$\Omega_{\rm b}/\Omega_{\rm m}$ instead of $\Omega_b$ does change
the contours somewhat.  For example in the origincal Cardassian model
the CMB contours expand to include higher values of $n$ but the 
$\Omega_{\rm m}$ range within the contours stays the same.

Parameters $\Omega_{\rm m}\in[0,1]$ and $n\in[0,2/3]$ 
have uniform priors and are chosen to cover the most interesting 
range of parameters. The typical grid size used in the parameter scan
was typically $(h,\Omega_m,n,q)\sim 5\times30^3$ (larger grids were also used
to scan larger parameter spaces than what is shown here).
For the Hubble parameter $h$ we use a Gaussian prior based on the
HST Key Project value $h=0.72\pm 0.08$ \cite{hst} 
and marginalize over $h$ in making all the plots unless otherwise stated
(we find that the choice of a flat or a Gaussian prior makes 
little difference).


\subsection{The original Cardassian model}

The original Cardassian model \cite{card} corresponds to the choice $q=1$,
leaving only three free parameters $(\omegam,h,n)$.
This model has been extensively studied by using SNIa data 
\cite{frith, gong, sen, nesseris, wang,zhu2}. 
The overall picture is that supernovae alone cannot constrain 
the parameter space particularly well and leave quite 
a large degeneracy 
in the $(\omegam,n)$ plane. CMB data has been utilized in \cite{sen, senwmap}, 
but only using the power spectrum peak positions. Finally, in \cite{zhu},
the original Cardassian scenario has been constrained by Sunyaev-Zeldovich/X-ray
data.

Here, we use the full shape of the WMAP power spectrum along with 
SNIa observations as well as consider the more general MPC model.
\begin{figure}
\begin{center}
{\centering
\mbox
{\psfig{figure=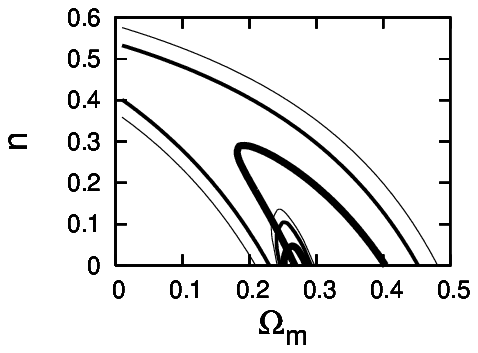,height=5cm,width=5cm}}
\mbox
{\psfig{figure=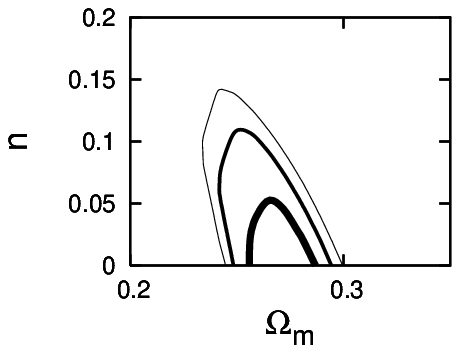,height=5cm,width=5cm}}
}
\caption{68, 95 and 99\% confidence contours for $\Omega_{\rm m}$ and 
$n$ for the original Cardassian model ($q=1$). Left: SNIa and CMB contours, 
right: combined contours}.
\label{fig:fig2}
\end{center}
\end{figure}
In Fig. \ref{fig:fig2} we show results for the original Cardassian model
from fitting to the SNIa and WMAP
data. We see that compared to the SNIa data, the WMAP data constrains
the model much more effectively. The combined contours demonstrate how 
the original Cardassian model is quite severely constrained by the current
data. The original Cardassian model seemingly does not
offer any advantage over the concordance \lcdm model.


\subsection{The MPC model}

\begin{figure}
\begin{center}
{\centering
\mbox
{\psfig{figure=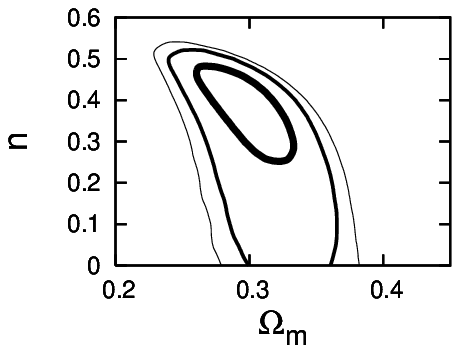,height=5cm,width=5cm}}
\mbox
{\psfig{figure=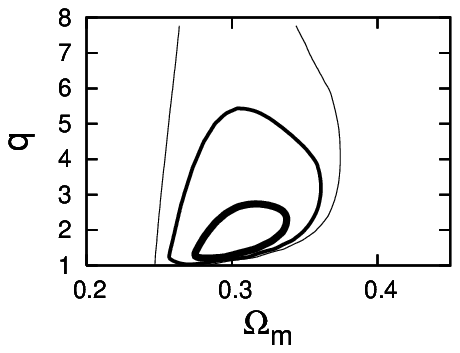,height=5cm,width=5cm}}
\mbox
{\psfig{figure=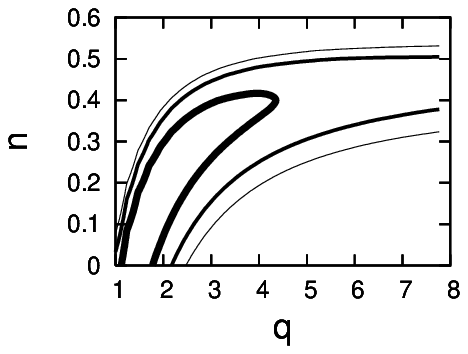,height=5cm,width=5cm}}
}
\caption{Joined 68, 95 and 99\% confidence contours for $\Omega_{\rm m}$ and $n$ 
(left), $\Omega_{\rm m}$ and $q$ (middle), and $q$ and $n$ (right), 
from fitting to SNIa and CMB data.  In each figure, we have marginalized the Hubble
parameter with a HST prior and the remaining parameters 
with a uniform prior.}
\label{fig:fig3}
\end{center}
\end{figure}
In the case for the more general MPC model, the situation is however somewhat
different. The combined fit to CMB+SNIa in the model is shown in 
Fig. \ref{fig:fig3} and the different sub figures show the confidence 
contours after marginalizing over one of the remaining parameters 
$\omegam,\ n,\ q$.
From the figures we see that the data prefers $\omegam\sim 0.3$, which
is expected in light of the tight constraints coming from the WMAP data.
Marginalizing over $\omegam$, we see that there is quite a large degeneracy in the 
$(q,n)$ plane. Comparing to previous works \cite{card2,savage} however, we see that
utilizing the full CMB spectrum along with SNIa data tightens the constraints
significantly. Even though in the full MPC model the data prefers a non-standard
model, there is no significant preference over the \lcdm model ($q=1,\ n=0$).

In a recent work \cite{koivisto}, the CMB TT power spectrum in the Cardassian
scenario was also considered and it was shown that the large scale part 
of the spectrum 
is modified, on par with what we find here and what we expect based from
analytical approximations \cite{mult3}.


\section{The Fluid Interpretation}

The Cardassian models can have different interpretations depending 
on the properties of the fluid. One can either consider that the modified
Friedmann equation arises from modified gravity, e.g., in extra dimensional
theories, or that gravity is standard but the fluid has unusual properties.
The fluid interpretation was introduced in \cite{card2} as a possible
scenario responsible for the modified Friedmann equation.

The basic idea of the fluid interpretation of the Cardassian models
is that gravity is four dimensional but the matter fluid has some pressure
which leads to a Friedmann equation that is effectively modified.
The starting point is the set of ordinary Einstein's equations,
\be{eins}
R_{\mu\nu}-\frac 12  g_{\mu\nu} {\cal R}=8\pi G T_{\mu\nu},
\ee
where the energy momentum tensor has the ordinary perfect fluid form
$T_{\mu\nu}={\rm diag}(\rho,-p,-p,-p)$. The Friedmann equation is of the 
normal form $H^2=(8\pi G)\rho/3$. The crucial point is that
now $\rho\neq\rhom$ but $\rho=\rho(\rhom)$ instead. Energy density must
be conserved, i.e., $T_{\mu\nu}^{;\mu}=0$, which leads to the usual
continuity equation for the fluid $\rho$:
\be{cont}
\dot{\rho}+3H(\rho+p)=0.
\ee
In order to have the modified Friedmann equation, Eq. (\ref{gencard}), 
we must make the identification $\rho=g(\rhom)$. Using this in the continuity
equation along with the fact that the matter fluid has no pressure,
gives an expression for the effective pressure of the fluid
\be{pres}
p=\rhom g'(\rhom)-g(\rhom),
\ee
where $'\equiv \partial/\partial\rhom$. It is then straightforward to calculate
the sound speed (in the adiabatic approximation) and the effective equation 
of state of the fluid:
\begin{eqnarray}
c_s^2 & = & \rhom{g''\over g'}\label{eq:sound}\\
\w & = & \rhom {g'\over g}-1\label{eq:w}.
\eea
Physically these properties require that the matter fluid must have
some interactions that lead to the pressure term of the form Eq. (\ref{pres})
(see \cite{card2} for a discussion).

Here we only consider the effects on the growth of large scale structure as it
is a good discriminator for models with a new fluctuating component.
The CMB spectrum was studied in \cite{koivisto}, where they find that
the large scale part of the spectrum can be modified radically in the MPC
scenario with fluctuations.


\subsection{Growth of perturbations: general case}
Since the effective fluid has non-zero pressure, perturbation
growth can be dramatically different compared to CDM.

For a single fluid, the fractional perturbation of the effective fluid
$\delta\equiv (\delta\rho)/\rho$, in momentum space
obeys \cite{lyth} (again in the adiabatic approximation, for modes for which
$k\geqslant H$)
\be{onefluid1}
\fl\ddot{\delta_k}+H\Big(2-3(2\w-c_s^2)\Big)
\dot{\delta_k}
-\frac 32 H^2\Big(1-6c_s^2+8\w
-3\w^2-\frac 23{c_s^2k^2\over a^2 H^2}\Big)\delta_k=0.
\ee

The physically interesting quantity is the perturbation
of the matter fluid, $\deltam$, which is related to $\delta$ by
\be{deltam}
\delta={(\delta\rho)\over\rho}=\rhom{g'\over g}\deltam=(1+w)\deltam,
\ee
from which it is clear that the two perturbations agree only when the fluid has
no pressure ($w=0$ or $g(\rhom)=\rhom$).

From Eq. \ref{onefluid1} one finds that the equation describing
the evolution of the matter perturbation is 
\bea{mfluid2}
\fl\ddot{\delta}_{\rm m,k} + H(2 - 3c_s^2)\dot{\delta}_{\rm m,k}
 + \Big\{-\frac 32 H^2\Big(1 -2c_s^2- 3c_s^4+4\w+9(c_s^2-\w)^2\Big)\nonumber\\
+ {c_s^2k^2\over a^2}+{\ddot{\w}\over 1+\w}\Big\}\delta_{\rm m,k}=0,
\eea
where we have made use of the identity 
\be{windet}
\dot{\w}=-3 H(1+\w)(c_s^2-\w).
\ee
With $c_s^2=0$, $\w=0$ (or $g(\rhom)=\rhom$) 
this reduces to the standard CDM expression,
\be{normalfluid}
\ddot{\delta}_{\rm m,k}+2H\dot{\delta}_{\rm m,k}-\frac 32 H^2\delta_{\rm m,k}=0.
\ee

\subsection{Growth of perturbations: MPC model}
Specializing to the MPC model where $g$ is given by Eq. (\ref{mpc}),
we find the following expressions:
\bea{MPCfluid}
\w & = & (n-1){\Omega_C a^{3q(1-n)}\over 1+\Omega_C a^{3q(1-n))}}\\
c_s^2 & = & \w\Big(1+{(n-1)q\over 1+n\Omega_C a^{3q(1-n))}}\Big),
\label{MPCcs2}
\eea
where have used the previously defined quantity $\Omega_C=(\omegam^{-q}-1)$.
From the expressions we see that in the fluid interpretation, the \lcdm model 
is reached in the limit $n=0,\ q=1$.

We have solved the equation describing the growth of the matter perturbation,
Eq. (\ref{mfluid2}), numerically in the MPC model. Following \cite{sandvik}, 
we use CMBFast to calculate the matter transfer function from an initial
scale free spectrum in the \lcdm model at $z=200$ and then evolve the 
perturbation according to Eq. (\ref{mfluid2}). The resulting 
matter power spectrum is shown
in Fig. \ref{Fig:cardmatter} for different parameter values
along with the SDSS\footnote{http://www.sdss.org} data from \cite{sdss}.
The figures are shown for the case $\omegam=0.27$, $h=0.71$, 
but the qualitative features of the spectrum are independent 
of the exact value of $\omegam$. The spectra are normalized to match
the matter power spectrum of the \lcdm concordance model at
$k=0.01\;h\,{\rm Mpc}^{-1}$.  

\begin{figure}
  \begin{center}
    \subfigure[]{\includegraphics[width=7.5cm]{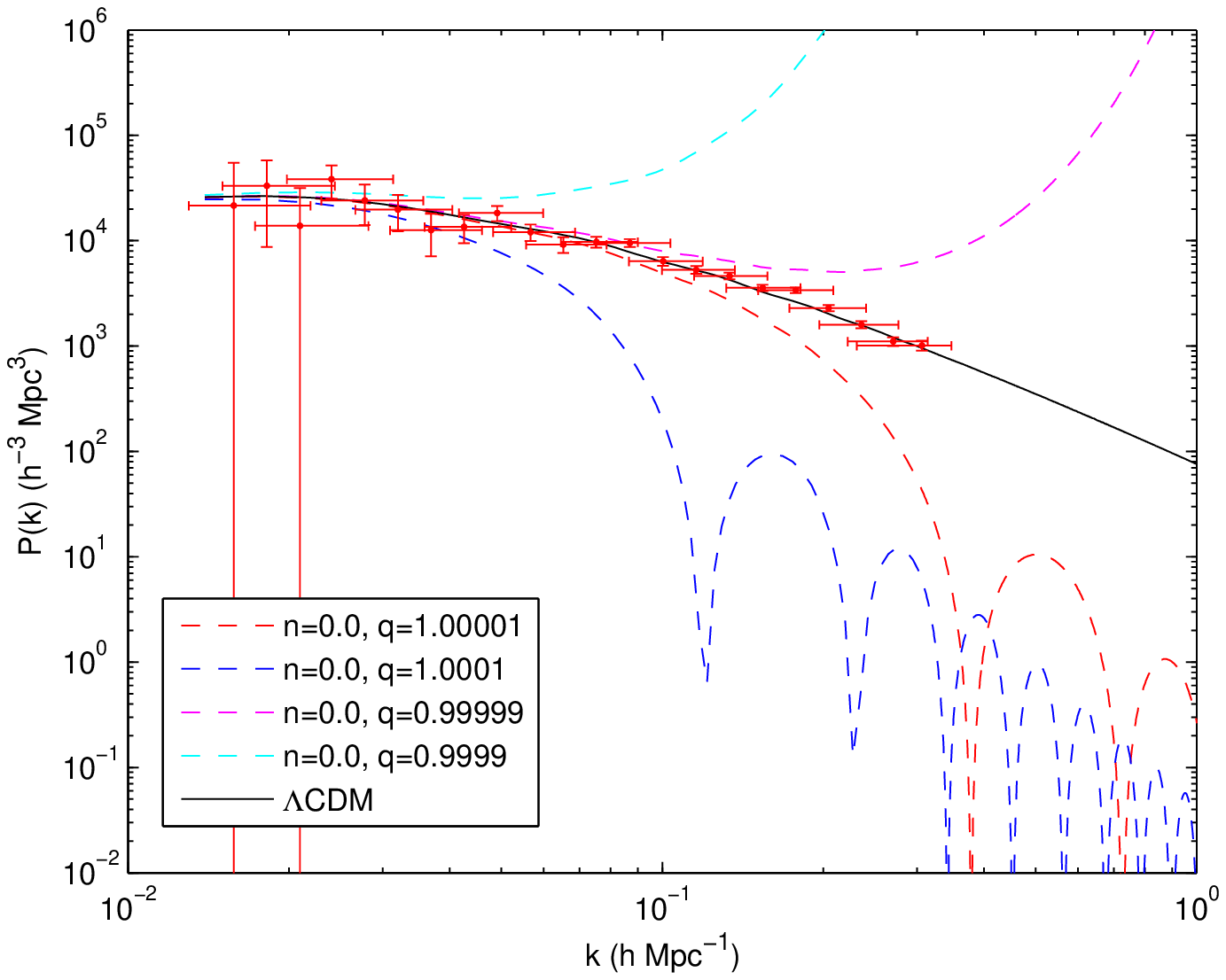}}
    \subfigure[]{\includegraphics[width=7.5cm]{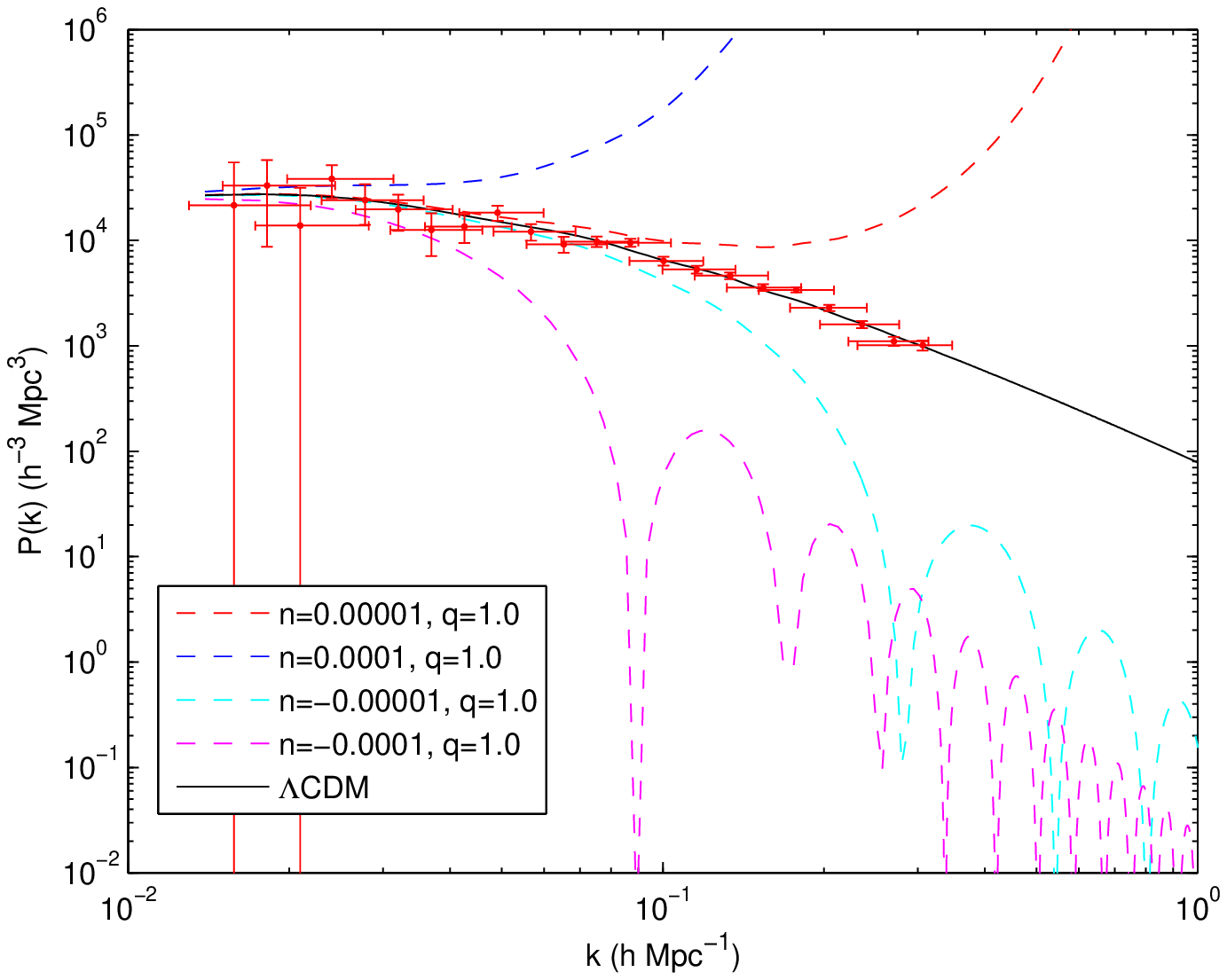}}
  \end{center}
  \caption{The matter power spectrum in the MPC model without baryons.
for small deviation of $q$ (a) and $n$ (b) from the
$\Lambda{}CDM$ values of $q=1,\,n=0$. Also shown are data points and 
error bars from the SDSS project.
}
  \label{Fig:cardmatter}
\end{figure}
From the figures it is clear that the matter power spectrum is 
drastically modified, much 
like in the Chaplygin gas scenario studied in \cite{sandvik}. Even a small
deviation from the \lcdm values ($n=0,\ q=1$) leads to either strong
oscillations or growth in the spectrum, depending on the signs of $n$
and $q-1$. This is explained by the non-zero sound speed that seeds
the non-standard evolution of $\delta_{m,k}$ due to the
$c_s^2k^2/a^2$-term in Eq. (\ref{mfluid2}). We know generally that
perturbations in a comsological fluid with $c_s^2>0$ will fluctuate
for scales smaller than the Jeans length, which is the scale when
pressure start to dominate over gravity. While for fluids with
$c_s^2<0$, they will be unstable and grow exponentially for the same
scales. A more thorough discussion on perturbations for fluids with
non-zero sound speed can be found in \cite{hu}.

%
%
%
%
%

Expanding the sound speed for the MPC model (\ref{MPCcs2})
in the small quantities $n$ and $q-1$, we find that they determine the
sign of $c_s^2$. For $n>0$ and $q-1\leq0$ $c_s^2$ will be negative
and hence give rise to exponentially growing perturbations, while for
$n\leq0$ and $q-1>0$ $c_s^2$ will be positive and the perturbations
will oscillate. This is exactly the behaviour we see in Fig.
\ref{Fig:cardmatter}. This behaviour for large $k$ can also be
understood directly from Eq. (\ref{mfluid2}). In the large $k$
limit we can ignore all but the $k^2$ term in the factor that
multiplies the zeroth derivative of the density perturbation. 
Then the differential equation simplifies to
\begin{equation} \label{mfluid2largek}
        \ddot{\delta}_{\rm m,k} + H(2 - 3c_s^2)\,\dot{\delta}_{\rm m,k} 
        + {c_s^2k^2\over a^2} \delta_{\rm m,k}=0\,
\end{equation}
We can get a feeling for how the density perturbation behaves in this
limit by looking at the case where the time variation of $a$ and
$c_s^2$ is relatively slow. Then both $a$ and $c_s^2$ will be
approximately constant, and the first derivative term in the
differential equation will vanish. The differential equation reduces
then to the common harmonic oscillator equation, with the following
solution
\begin{equation} \label{largeksol}
	\delta_{m,k}(t_0)\sim e^{- ic_skt_0}.
\end{equation}
From this we see that a positive $c_s^2$ will give oscillating
perturbations, while a negative $c_s^2$ gives exponentially growing
perturbations. We expect the perturbations to exhibit such behaviour
for scales smaller than the Jeans length. The Jeans length is given by
the expression
\begin{equation} \label{defJeans}
	\lambda_J^2\equiv|c_s^2|\frac{\pi}{G\rho}\,.
\end{equation}
For the Cardassian model the Jeans length today can be written as
\begin{equation} \label{cardJeans}
	\lambda_{J,0}^2=\frac{8\pi^2}{3}\frac{n-1}{H_0^2}
        \frac{\Omega_C}{1+\Omega_C}\left(1+\frac{(n-1)q}{1+n\Omega_C}
        \right).
\end{equation}
The pressure starts to dominate when
\begin{equation} \label{defkeq}
	k>\frac{2\pi a}{\lambda_J}\equiv{k_J}
\end{equation}
Let us look explicitly at the Cardassian model with parameters $n=0$
and $q=1.00001$. Using the expression for the Jeans length in Eq.
(\ref{cardJeans}), we find that $k_J\approx0.6\,$h/Mpc. According to
our simplified analysis, the perturbations should then start to grow
exponentially when $k>0.6\,$h/Mpc. This agrees fairly well with what
we see in Fig. \ref{Fig:cardmatter}a.

In \cite{beca} it was argued that the very tight constraints in the
Chaplygin case arising from the strong deviations in the matter power
spectrum could be relaxed by considering the effect of baryons. Even
though the constraints are relaxed slightly, the overall effect is very
strong and the fluctuating Chaplygin gas scenario remains tightly
constrained \cite{sandvik}, even without considering the overall 
normalization of the spectrum, which tightens the constraints even more.
Here, we expect a similar situation to apply
but for completeness consider the effect of adding the baryons to the
system.

For $k\geq H$ the evolution of linear density perturbations of the
Cardassian and the baryon fluids are given approximately by a coupled
set of second order differential equations. Taking the equation of
state parameter and the adiabatic sound speed of matter to be zero,
these differential equations can be written as (\cite{lyth,sandvik}):
\bea{onefluid2}
\fl \ddot{\delta_k}+ H\Big(2-3(2\w-c_s^2)\Big)
\dot{\delta_k}- \Big(\frac{3}{2}H^2(7w-3w^2-6c_s^2)
-(\frac{c_sk}{a})^2\Big)\delta_k\nonumber\\
-\frac{3}{2}H^2(1+w)\frac{\rho\delta_k+\rho_b\delta_{b,k}}{\rho+\rho_b}=0
\eea
and
\begin{equation} \label{EQ:diffEqBaryon}
	\ddot{\delta_b}+2H\dot{\delta_b}-\frac{3}{2}H^2
	\frac{\rho\, \delta+\rho_b\delta_b}{\rho+\rho_b}=0\,,
\end{equation}
where $\delta$ is again the density perturbation of the Cardassian fluid,
$\delta_b$ the density perturbation of the baryon fluid and $w$ and
$c_s^2$ are the equation of state parameter and adiabatic sound speed
of the Cardassian fluid.

We solve the equations as before and the resulting matter power
spectrum is shown in Fig. \ref{Fig:baryon}.
We see that adding the baryons to the system does improve the situation
as the oscillatory behaviour disappears, and the main effect is a 
reduction in small-scale power.  The constraints are still very strong, 
but bearing in mind that we have neglected other effects which can 
result in less power on small scales, e.g. neutrino masses and 
a scale-dependent bias between matter and galaxy clustering, this 
simple analysis cannot rule out these models completely.  However, 
as shown in \cite{koivisto}, the integrated Sachs-Wolfe effect in 
the CMB leads to very strong constraints on models with $c_s^2 > 0$.  
In the cases where $c_s^2 < 0$, 
the exponential blow-up is still present, and is strongly disfavoured 
by the galaxy power spectrum.  
\begin{figure}
  \begin{center}
    \subfigure[]{\includegraphics[width=7.5cm]{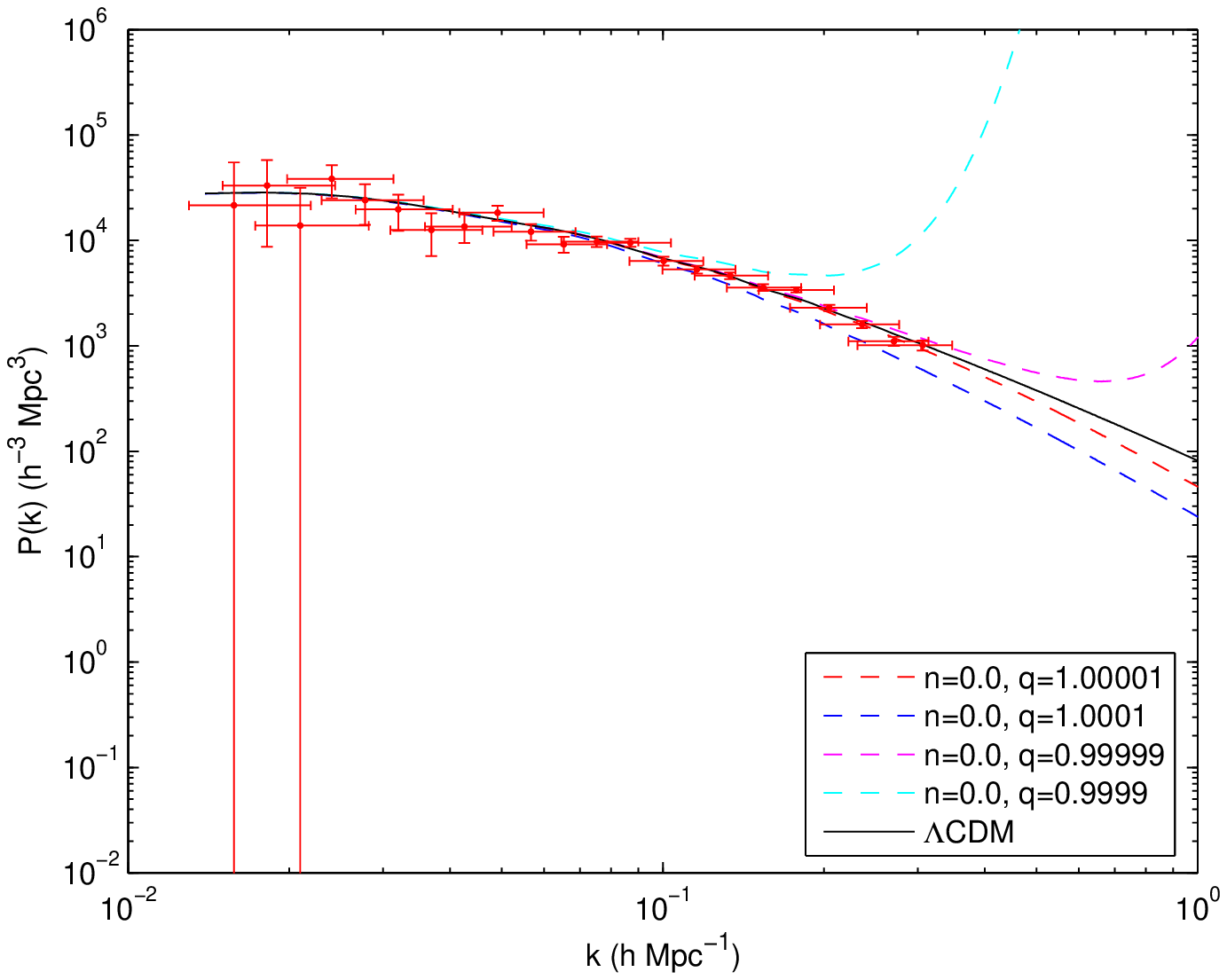}}
    \subfigure[]{\includegraphics[width=7.5cm]{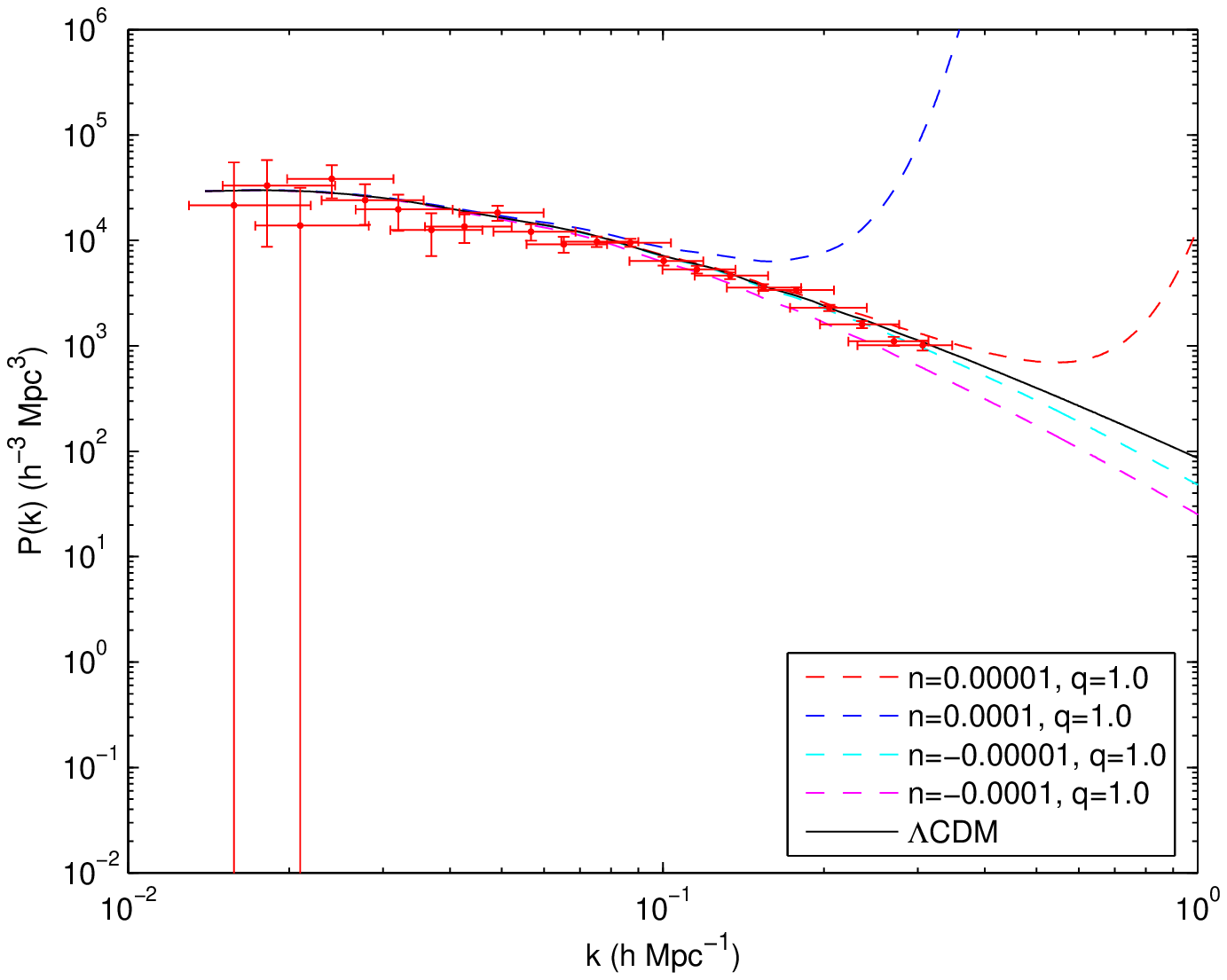}}
  \end{center}
  \caption{The matter power spectrum in the MPC model with baryons
for small deviation of $q$ (a) and $n$ (b) from the
$\Lambda{}CDM$ values of $q=1,\,n=0$.
The oscillatory behaviour has now vanished for $n<0$ and $q>1$ but 
still very small deviations from the \lcdm model worsen the fit to 
SDSS data.}
  \label{Fig:baryon}
\end{figure}

Here, considering the effect on $\sigma_8$, does not tighten the
constraints like in the Chaplygin gas case.
This can be seen by comparing how the linear growth
is changed in the two scenarios: the Chaplygin gas case is shown in 
\cite{mult2} and the MPC case in \cite{multam}. Linear growth
in the Chaplygin gas scenario changes radically for very small deviations
from the parameter values corresponding to the \lcdm model, whereas
for the MPC model this is not the case. We have checked  that this also
holds for the parameter values considered here
by calculating the linear growth in the MPC scenario for $n<0,\ q<1$.
Linear growth is modified, but not as radically as in a Chaplygin gas 
dominated universe and the LSS constraints are much tighter.
Therefore adding baryons gives somewhat more freedom
in choosing the parameters in the MPC scenario, but one should 
bear in mind the strong constraints from the ISW effect \cite{koivisto}. 
As is clear from Fig.
\ref{Fig:baryon}, the parameter space is highly constrained and  
the model offers no good alternative to the \lcdm model.


\section{Conclusions}

We have considered the observational constraints arising from current
CMB, SNIa and LSS data on commonly considered models within the Cardassian
framework, the original Cardassian model and the Modified Polytropic
Cardassian model. Both models were introduced as an alternative explanation
to the acceleration of the universe.

Considering that the models only have an effect on the background evolution,
CMB and SNIa observations constrain the allowed parameter space. We find that,
within the approximation that only background evolution is modified,
using the full CMB spectrum constrains the original Cardassian model significantly
and that the model offers no advantage over the \lcdm model. Adding a new 
degree of freedom by extending the consideration 
to the Modified Polytropic
Cardassian model relaxes the constraints somewhat but still the parameter space
is more constrained than what one finds using the supernovae data only.
The constraints indicate that a somewhat non-standard model is slightly preferred
over the \lcdm model but not significantly.

Viewing the Cardassian models as arising from a interacting fluid constrains
the scenario critically. Considering the growth of fluctuations along with the
SDSS data, we find that any deviation from the \lcdm model is strongly disfavoured,
with or without including baryons in the calculation. The fluid interpretation
of the Cardassian model is hence effectively excluded as a viable alternative
to the \lcdm model. This is further supported by recent work \cite{koivisto}, 
where the CMB TT power spectrum in the Cardassian model with fluctuations is
demonstrated to be modified strongly on large scales.

In short, even small deviations from the \lcdm model 
within the fluid interpretation of the Cardassian model are strongly disfavoured by current
data. If on the other hand, only background evolution is modified, the data 
indicates a slight preference over the \lcdm model but the difference is not significant.

\label{conclusions}


\ack
\O E thanks NORDITA, where 
parts of this work were carried out, for their hospitality.    
MA and \O E acknowledge support from the Research Council of 
Norway through grant 159637/V30.


\end{document}